\documentstyle[psfig]{mn}
\title[Isolated Black Holes]{X-Rays from Isolated Black Holes in the Milky Way}

\author[E. Agol and M. Kamionkowski]{
Eric Agol\thanks{Chandra Fellow, Email: agol@tapir.caltech.edu} and 
Marc Kamionkowski \\
California Institute of Technology, Mail Code 130-33, Pasadena, CA 91125 USA}

\onecolumn
\date{August 2001}
\pagerange{\pageref{firstpage}-\pageref{lastpage}}
\pubyear{2001}

\begin{document}

\maketitle

\label{firstpage}

\begin{abstract}
Galactic stellar-population-synthesis models, chemical-enrichment models, and
long-duration Bulge microlensing events indicate about $N_{\rm
tot}=10^8-10^9$
stellar-mass black holes reside in our Galaxy.  We study X-ray emission 
{}from accretion from the interstellar medium on to isolated black holes.
Although isolated black holes may be fewer in number than neutron stars,
$N_{NS} \sim 10^9$,
their higher masses, $\langle M\rangle \sim 9 {\rm M}_\odot$,  and smaller space 
velocities, $\sigma_v \sim 40$ km~s$^{-1}$ result in Bondi-Hoyle accretion rates 
$\sim 4 \times 10^3$ times higher than for neutron stars.  
We estimate that $\sim 10^4$ isolated black holes within the Milky Way should 
accrete at $\dot M > 10^{15}$ g~s$^{-1}$, comparable to accretion rates inferred for 
black-hole X-ray binaries given a total number of black holes
$N_{\rm tot}=N_9 10^9$.
If black holes accrete at efficiencies only $\sim
10^{-4}(N_{NS}/N_{\rm tot})^{0.8}$
of the neutron-star accretion efficiency, a comparable number of each
may be detectable.  We make predictions for the number of isolated
accreting black holes in our Galaxy which can be detected with X-ray surveys
as a function of efficiency, concluding that all-sky surveys at a depth of 
$F= F_{-15}10^{-15}$ erg~cm$^{-2}$~s$^{-1}$~dex$^{-1}$ can find $N(>F)\sim 10^4 N_9
(F_{-15}/\epsilon_{-5})^{-1.2}$ isolated accreting black holes for a velocity 
dispersion of 40 km~s$^{-1}$ and X-ray accretion efficiency of 
$\epsilon = \epsilon_{-5} 10^{-5}$.  Deeper surveys of the Galactic plane 
with {\em Chandra} or {\em XMM-Newton} may find tens of these objects per
year, depending on
the efficiency.  We argue that a minimum mass can be derived for microlensing
black-hole candidates if they are detected in the X-ray.
\end{abstract}

\begin{keywords}
accretion, accretion discs --- black hole physics --- Galaxy: stellar content 
--- X-rays: ISM, stars
\end{keywords}

\section{Introduction}

A black hole may only be detected via its interactions with surrounding 
matter or light.  Black holes in binary-star systems are strong
and transient X-ray sources
fed by accretion from a stellar companion.  However, isolated black 
holes---that is, black holes without companions or black holes in wide binaries---can 
easily escape detection.  These may constitute the majority of the population
of black holes since the fraction in close binaries with $a < 100$ AU is $\sim$ 0.01
based stellar evolution calculations for massive star binaries
(Chris Fryer, priv. comm. 2001).  Young isolated neutron stars are much easier to 
detect as pulsars, but once the magnetic field, spin, and thermal energy decay 
away, an old neutron star becomes as invisible as a black hole.  Several
authors (Ostriker, Rees \& Silk 1970,  Treves \& Colpi 1991, Blaes \&
Madau 1993) have proposed searching for isolated old neutron stars lit
up by accreting matter from the interstellar medium (ISM).   A tenth of
the rest-mass energy of the accreting gas can be directly converted to
luminosity assuming that all the accreted material is thermalised as it 
hits the neutron-star surface.   These predictions motivated searches that
turned up several isolated neutron-star candidates (Stocke et al. 1995,
Walter et al. 1996, Wang 1997, Schwope et al. 1999); however, it is possible 
that these are young cooling neutron stars powered by processes other than 
accretion (Treves et al. 2000).

The known Galactic black-hole X-ray-binary population
has a velocity dispersion smaller by a factor of $\sim 5$ than neutron 
stars (Hansen \& Phinney 1997, White \& van Paradijs 1996, van Paradijs \&
White 1995) and average 
mass larger by a factor of $\sim 6$ (see below and Bailyn et al. 1998, Thorsett \& 
Chakrabarty 1999).  The accretion rate for an object of mass $M$ accreting from
the interstellar medium is determined by the Bondi-Hoyle formula 
(Bondi \& Hoyle 1944),
\begin{equation}\label{bondi}
\dot M = {\lambda 4 \pi G^2 M^2 n \mu \over (v^2 + c_s^2)^{3/2}},
\end{equation}
where $v$ is the velocity of the black hole with respect to the
local interstellar medium, $n$ is the number density of hydrogen in the 
interstellar medium, $\mu = \rho/n$ where $\rho$ is the particle mass 
density,
$c_s$ is the sound speed of the interstellar medium, and $\lambda$ is a
parameter of order unity (we set $\lambda =1$ hereafter).   Assuming that
isolated black holes have the same mass range and velocity dispersion as 
black holes in X-ray binaries, one would expect the typical accretion rate 
on to isolated black holes from the ISM to be larger by a factor of 
$\sim 4\times 10^3$ compared to neutron stars.   The expected detection 
rate of isolated black holes accreting from the ISM may be reduced as the
number of black holes in our Galaxy may be smaller by a factor of $\sim 10$
than the number of neutron stars, and the lack of a hard surface may
result in lower luminosities for black holes accreting at small accretion
rates.  Garcia et al. (2001) argue that black holes have quiescent 
X-ray efficiencies $\sim 10^{-2}$ of neutron stars; however, Bildsten \&
Rutledge (2000) argue that the detected X-rays may have nothing to do with 
accretion, and V404 Cyg is an exception to this rule indicating that there 
may be a wide range of accretion efficiencies.  If black holes do have a
reduced efficiency compared to neutron stars, this reduction can be 
compensated by the larger accretion rates and the smaller scale height 
of black holes which increases their number density in the mid-plane where 
most of the interstellar gas mass is located.  Black 
holes do not have an intrinsic magnetic field, so accretion can proceed 
uninhibited by magnetic pressure or torque.  Thus, we are motivated to compute 
the expected numbers and accretion rates of isolated black holes in our Galaxy.

The problem of black holes accreting from the interstellar medium has
been considered by several authors.  We approach the problem in a different
way from previous authors (listed below) by using the properties of black-hole 
X-ray binaries and black-hole microlensing candidates to constrain the 
phase-space distribution of black holes within the entire Galactic disc 
(discussed in detail in Agol \& Kamionkowski 2001, in preparation), 
while considering the properties of all phases of the ISM, deriving the 
distribution of accretion rates including
radiative pre-heating, and finally deriving the distribution of X-ray 
fluxes.  Grindlay (1978) considered spherical-accretion models for 
X-ray sources now known to be accreting from companions.
Carr (1979) estimated the luminosity of black holes
in the galactic disc, but did not predict their detection probability.
McDowell (1985) computed the visual and infrared fluxes of black 
holes accreting spherically from molecular clouds, but only out to 150 pc.  
Campana \& Pardi (1993) computed the fluxes of black holes within the 
disc of the Galaxy accreting spherically from molecular clouds, without
computing the probability distribution of accretion rates.
Heckler \& Kolb (1996) computed the fluxes of a putative halo population of black 
holes accreting spherically while passing within 10 pc of the Sun.
The large velocities, low densities, and large scale heights of a halo
population greatly reduces the number of black holes at a given accretion 
rate.  Popov \& Prokhorov (1998) took into account the spatial distribution 
of the warm interstellar medium, assuming the black holes receive a kick
velocity at birth of $\sim 200$ km~s$^{-1}$, which greatly reduces the accretion
rates.  Fujita et al. (1998) computed the fluxes of black holes in the 
disc accreting via an advection-dominated accretion flow at a distance
of 400 pc (the Orion cloud) for a few velocities, without computing the 
full probability distribution.  Grindlay et al. (2001) estimate that
of order $10^3$ black holes in molecular clouds might be detectable by
the proposed Energetic X-ray Imaging Survey Telescope ({\em EXIST}) experiment
based on an accretion efficiency of $10^{-2}$.

The spectrum of an accreting black hole has been computed in the
spherical-accretion limit (Ipser \& Price 1977, 1982, 1983); however,
as we show below, the accreted material will possess non-zero angular
momentum, possibly forming an accretion disc which may increase
the efficiency of accretion.  Thus, we parameterize our results in
terms of the unknown accretion efficiency.

In section 2 we summarize the known properties of binary and isolated
black holes and their distribution in phase space.  In section 3 we estimate 
the accreted angular momentum, discuss the range of accretion efficiencies,
and discuss the effects of radiative feedback.  In section 4 we discuss 
the properties of the interstellar medium necessary for computing the 
accretion rates of black holes.  In section 5 we compute the distribution 
of accretion rates and luminosities for black holes within the disc of 
the galaxy.  In section 6 we estimate the detection rates of black holes 
under the persistent and transient assumptions, and then we
predict the fluxes as a function of distance for the three candidate
black-hole microlenses.  Finally in section 7 we summarize.

\section{Isolated black holes in our Galaxy}

The relevant results from Agol \& Kamionkowski (2001) on the number and
distribution of black holes within the Galaxy are summarized here.

\subsection{Number of black holes}

Three microlensing events toward the Galactic Bulge have been found
by the MACHO and OGLE teams (Bennett et al. 2001, Mao et al. 2001).
These events are of long enough duration ($>$ 1 year) to observe variations 
in the microlensing light curve due to the motion of the Earth around the
Sun, an effect known as microlensing parallax.  This breaks some
of the degeneracy between distance, velocity, and mass of the lensing
object;  when combined with a model for the velocity and number density
of black holes in the Galaxy, this results in an estimate of the likely
mass of the lensing objects.   In all three cases, the objects likely
have masses greater than 3M$_\odot$, while they are not detected as
main-sequence stars, leading to the conclusion that they are most
likely black holes (Bennett et al. 2001, Agol \& Kamionkowski 2001).
Based on the probability of microlensing toward the Bulge and assuming
that isolated black holes have the same scale height as black holes
in binaries, one concludes that the total number of black holes in the 
Milky Way disk is %$N_{\rm tot} = 1_{-0.3}^{+0.8}\times 10^9(f/0.2)^{-1}$ 
%(1-$\sigma$ errors) 
%the number of black holes is 
$2\times 10^8 < N_{\rm tot}(f/0.2)
< 4\times 10^9$ at 95 per cent confidence,
for a microlensing detection efficiency $f$, 
estimated to be 20 per cent (Alcock et al. 2000).   
Given the uncertainty in these estimates, we will make predictions
based on $N_{\rm tot}=N_910^9$ black holes, and note that our results 
scale with $N_9$. 

As discussed below, there is some evidence that the velocity
dispersion of isolated black holes is similar to that of the
stellar disk. We thus assume that the number density of black
holes is distributed as 
\begin{equation}
N_{\rm BH} = N_\odot \exp{\left[-{|Z|\over H} -{(R-R_\odot)\over R_{\rm disc}}\right]},
\end{equation}
where $N_\odot$ is the number density at the solar circle, $Z$ and $R$
are the cylindrical Galactic coordinates, $R_\odot = 8$ kpc is the 
distance to Galactic centre, $H$ and $R_{\rm disc}$ are the disc scale height 
and scale length.  We will assume that the black-hole disc scale height is
$H= 375$ pc and scale length is $R_{\rm disc}=3$ kpc.  This implies a local
number density of $N_\odot = 1.64\times10^5 N_9 (H/375 {\rm pc})^{-1}$.

\subsection{Velocity distribution of isolated black holes}

White \& van Paradijs (1996) have estimated that the rms distance from the 
Galactic plane for black-hole X-ray binaries, 
$z_{\rm rms}$, is about 450 pc, which corresponds to a scale height of
$H \sim 320$ pc and velocity dispersion of
$\sigma_v \sim 40$ km~s$^{-1}$. 
This is consistent with the radial-velocity measurements of
low-mass black-hole X-ray binaries when corrected for Galactic rotation
(Brandt et al. 1995, Nelemans et al. 1999), with the exception
of GRS 1655-40 which has $v_r = -114$ km~s$^{-1}$.
We have improved on the work by White \& van Paradijs (1996) by 
using a maximum-likelihood analysis based on 19 black-hole X-ray binaries,
concluding that $H = 375_{-60}^{+135}$ pc (Agol \& Kamionkowski 2001).
In addition, the three isolated black holes have a vertical velocity 
dispersion of $\sigma_z< 27$ km~s$^{-1}$ at 95 per cent confidence, indicating that their
kinematics, and thus spatial distribution, is similar to that of
black holes in X-ray binaries.  Thus, we take the isolated black holes to
have the same velocity dispersion and scale height as black hole
X-ray binaries.

\subsection{Mass distribution of isolated black holes}

The distribution of masses of black holes in X-ray binaries and
microlensing events can be fit with a power law mass distribution,
\begin{equation}\label{dndm}
dN/dM=n_M M^{-\gamma},
\end{equation}
with a lower mass cutoff at $M_1$ and upper mass cutoff of $M_2$.  We obtain
a best fit for $\gamma=0.6$, $M_1=4 {\rm M}_\odot$, and $M_2=14
{\rm M}_\odot$ (Agol \& Kamionkowski 2001).
Since there may be larger-mass black holes that have not yet been detected,
we impose an upper limit mass cutoff of $M_2=50 {\rm M}_\odot$ 
based on the theoretical mass limit for pulsational stability (Schwarzchild \&
H\"arm 1959), assuming that about half of the mass is lost before collapse.
This yields $M_1=4{\rm M}_\odot$ and $\gamma=2.4$, close to the value of the Salpeter IMF
(Agol \& Kamionkowski 2001).
The average mass of either distribution is  $\sim 9 {\rm M}_\odot$, close to the 
mean of the data $\langle M\rangle =9.6M_\odot$.  We have normalized this 
to unity, $n_M=(\gamma-1)/(M_1^{1-\gamma} -M_2^{1-\gamma})$, and we assume 
that it applies throughout the Galaxy.  We will later mention how our
predictions differ for $M_2=13{\rm M}_\odot$, but will otherwise use $M_2=50{\rm M}_\odot$.

\section{Efficiency of accretion}

\subsection{Angular momentum of accreted material}

The Bondi-Hoyle formula may be appropriate for computing the accretion rate, 
but not the efficiency.  If the accreting gas were perfectly uniform,
then the accreted angular momentum is quite small, and the accretion
flow follows a one-dimensional supersonic solution, for which the
luminosity can be predicted (Shapiro \& Teukolsky 1983, Ipser \& Price
1977).  However,
small perturbations in the density or velocity of the accreting gas 
will lead to an angular momentum large enough to circularise gas before 
it accretes (Shapiro \& Lightman 1976).  If there is a gradient in the 
density field that is perpendicular to the direction of motion of the 
black hole, then the accreted angular momentum scales as 
\begin{equation}
l= {1\over 4} {\Delta \rho \over \rho} v r_A,
\end{equation}
where $\Delta \rho$ is the difference in velocity between the top
and bottom of the accretion cylinder and $r_A=GM/(v^2+c_s^2)$ is the accretion
radius.
Numerical simulations confirm that this formula 
is correct to order of magnitude (Ruffert 1999).  
The observed density fluctuations in our galaxy scale
as $\delta \rho /\rho \sim (L/10^{18} {\rm cm})^{1/3}$ (Armstrong, Rickett,
\& Spangler 1995), extending down to a scale of $\sim 10^{8}$ cm.  The exponent 
of this scaling is nearly consistent with a Kolmogorov spectrum of 
density (Dubinski et al. 1995).  
Evaluating this at $L=2 r_A$, we can
then find the radius of the resulting accretion disc, $r_{\rm disc}$,
by equating the angular momentum of the gas with the Keplerian
angular momentum due to the black hole,  $l_{\rm Kep} = 
\sqrt{GMr_{\rm disc}}$.  This gives
\begin{equation}
{r_{\rm disc}\over r_g} = 10^{4} r_g \left({M\over9{\rm M}_\odot}\right)^{2/3} \left({\sqrt{v^2+c_s^2}
\over 40 {\rm km~s^{-1}}}\right)^{-10/3}.
\end{equation}
So, we see that a disc will almost always form in interstellar accretion.  
A similar argument is given in Fujita et al. (1998) for the case of
molecular clouds, for which the slope of velocity fluctuations is similar
to that expected by a Kolmogorov spectrum (Larson 1981). The neutral 
interstellar medium seems to follow a power-law with a similar slope 
(Lazarian \& Pogosyan 2000), which will also lead to non-zero accreted 
angular momentum;  however, the length scales probed in molecular and 
atomic gas are much larger than the scale of the accretion radius.

The accretion time scale for the disk scales as $t_{d} =
\alpha_{SS}^{-1}(GM/r_{\rm disc}^3)^{-1/2}
(r_{\rm disc}/h)^2 $ where $h$ is the height of the disc and $\alpha_{SS}$
is the ratio of the viscous stress to the pressure in the disk.
If the velocity and density fields of the ISM are incoherent on
the scale of these fluctuations, then the angular momentum can change on 
a time scale of $t_a=r_A/v$ as gas will be accreted with the angular momentum 
vector pointing in a different direction after this time scale.  Thus, the 
angular momentum of the disc can be reduced due to a random walk of the accreted 
angular momentum vector if $t_a \ll t_d$.  We estimate
\begin{equation}
{t_d\over t_a} \sim 2 {r_A\over 2\times 10^{13} {\rm cm}}
\left({0.1 r \over h}\right)^2 \left({\alpha_{SS}\over 0.1}\right)^{-1},
\end{equation}
which indicates that there may only be a slight reduction in the angular
momentum since these time scales are comparable.
The residual angular momentum means that we {\em cannot}
use spherical-accretion formulae for the luminosity, but we must 
estimate the efficiency of the accretion disc that forms.

\subsection{Range of efficiencies}

We assume that the black holes radiate a spectrum with spectral index 
$\alpha=1$, where $F_\nu \propto \nu^{-\alpha}$, consistent with the
typical spectrum of black-hole X-ray binaries, so that equal energy is
emitted per unit decade of energy.
We define $L_{\rm ion}$ as the luminosity from 1 to 1000 Ryd (13.6 eV),
so the efficiency is 
$\epsilon_{\rm ion}= L_{\rm ion}/(\dot M c^2)$.  
Estimates of the accretion rate in Cygnus X-1 indicate that it is
accreting with an X-ray efficiency of $\epsilon_{\rm ion}\sim 10^{-1}$ (Shapiro \& Teukolsky
1983).   However, at low accretion rates the density of the gas decreases,
so it may not be able to cool,
reducing the radiative efficiency.  By estimating the mass accretion
rates in black-hole X-ray binaries and X-ray luminosities in the
quiescent state, Lasota (2000) concludes that the 2--10 keV efficiency is
about $2\times 10^{-6}$ assuming that this luminosity
is solely due to accretion, implying $\epsilon_{\rm ion} \sim 10^{-5}$.  
Low upper limits on the X-ray Bondi-accretion efficiencies are inferred for
some supermassive black holes in elliptical galaxies (Loewenstein et al. 2001).
The Galactic-centre black hole has an X-ray efficiency of $\sim 4\times
10^{-8} - 2\times10^{-6}$ in the 2--10 keV band for a Bondi-Hoyle accretion 
rate of $10^{-6} {\rm M}_\odot$~yr$^{-1}$ (Baganoff et al. 2001).  Thus, we parameterize 
our results in terms of an unknown X-ray efficiency.

For certain ranges of parameters, the disc may be unstable to the 
hydrogen-ionisation disc instability (Lasota 2001), creating intermittent 
outbursts of higher luminosity between long periods of quiescence, with 
changes in luminosity of $10^{5}$.  We will discuss the possible effect 
this may have on the number of detectable sources in section 6.2.

\subsection{Radiative feedback}

The gas in the vicinity of the black hole will be heated by the
ionising radiation field produced by accretion.   If the temperature
rises enough, then the accretion radius, and thus the accretion rate,
will decrease (Shvartsman 1971, Ostriker et al. 1976).   For an
$\alpha=1$ spectrum, the mean opacity from 1-1000 Ryd of neutral gas with cosmic
abundance is $\sigma_{\rm cold}=3.87\times 10^{-18}$ cm$^{2}$ per hydrogen 
atom.  The heating rate is then given by $H=L_xn_H\sigma_{\rm cold}
/(4\pi R^2)$, where $R$ is the distance from the black hole.  For 
temperatures smaller than $\sim 10^4$ K and for a 
small ratio of radiation energy density to gas energy density, 
the heating rate greatly exceeds the cooling rate.   Consider a fluid element 
directly in front of the path of a moving black hole.  As the black hole 
approaches with velocity $v$, the ratio of the total heating rate to the 
initial energy density, $e_0$ of the gas is
\begin{equation}
{e_{\rm tot}\over e_0}={1\over e_0}\int_R^\infty dR H/v = 2\times 10^5 
\epsilon_{\rm ion} v_{40}^{-2} (r/r_A) T_4^{-1},
\end{equation}
where $e_{\rm tot}$ is the total energy density absorbed by a fluid element
(neglecting cooling), $v=v_{40} 40$km~s$^{-1}$ and $T=T_4 10^4$ K.  Thus, even 
for small efficiencies
the gas is strongly heated outside the accretion radius.  Once H and He 
are ionized, the mean opacity drops by three orders of magnitude, and the 
gas remains at around $2\times 10^{4}$ K.
This has been confirmed with detailed numerical simulations by
Blaes, Warren \& Madau (1995).   Above this temperature, the
cooling and heating times become much longer than the accretion time
so the gas falls in adiabatically.
Thus we take the temperature of the accreted gas to be $T_0=2\times 10^4$ K
for all phases of the ISM except the hot HII.

\section{Gas in our Galaxy}

\subsection{Phases of the ISM}
The interstellar medium in our Galaxy consists of at least five identifiable 
phases (Bland-Hawthorn \& Reynolds 2000) in approximate
pressure equilibrium, $P \sim 2\times 10^3 k_B$.  
The highest accretion rates will occur in the regions
of the highest density.
This means that the observable sources will be dominated 
by the giant molecular clouds (GMCs) consisting mostly of H$_2$ ($T \sim
10$ K) and the cold neutral medium consisting of HI clouds
(CNM, $T \sim 10^2$ K).  These phases comprise approximately 
40 and 50 per cent of the total mass of gas in our Galaxy.

The warm neutral and ionised components ($T\sim 8000$K) of the 
interstellar medium have densities that are $\sim 10^{-2}$ of 
the CNM, so the maximum accretion rates are $\sim 10^{-2}$ times
smaller than than the maximum for the CNM, while
the hot component of the ISM ($T\sim 10^6$K) has a density $\sim 10^{-2}$ of the
warm components, reducing the maximum accretion rate by another
factor of $10^{-2}$.  However, these phases fill a much larger
volume in the interstellar medium, and thus should be considered.

We next refine the above estimates by considering the distribution
of each phase as a function of number density.

\subsection{Filling fraction of interstellar medium}

To estimate the distribution of accretion rates for black holes
in our Galaxy requires knowing the volume of the Galaxy filled
by gas of a given number density.  We define this as 
\begin{equation}
{df \over dn} = f_0 n^{-\beta},
\end{equation}
where $df/dn$ is the fraction of volume filled by gas with number
density between $n$ and $n+dn$, and $f_0$ is a normalization constant.
Berkhuijsen (1999) has shown that
$\beta_{GMC} \sim 2.8$, while $\beta_{CNM} \sim 3.8$.  This can 
be seen for the molecular clouds by noting that the mass function 
of molecular clouds in our Galaxy scales as $dN/dM \propto M^{-1.6}$
(Dickey \& Garwood 1989), while number density of clouds scales
with size, $D$, as $n \propto D^{-1}$ (Larson 1981, Scoville
\& Sanders 1987).
Assuming that clouds have a spherical volume, one finds 
$df/dn \propto V(dM/dn)(dN/dM) \propto n^{-2.8}$.   Since the
number density-size relation is derived for the average density
of clouds, $\beta$ may be slightly smaller since each cloud may
contain regions where the gas density is higher than average.

We assume that the GMCs span number densities
{}from $n_1=10^2$ to $n_2=10^5$ cm$^{-3}$, while the CNM spans number
densities from $n_1=10$ to $n_2=10^2$ cm$^{-3}$.  
Finally, we assume that the scale height of
the molecular clouds is 75 pc, while the scale height of the CNM
is 150 pc.  

The surface mass density of molecular gas as a function of radius we
take from Clemens, Sanders \& Scoville (1988) and Scoville \& Sanders 
(1987), while the 
distribution of the CNM we assume is constant with $\Sigma = 4.5 
{\rm M}_\odot~$pc$^{-2}$ for $R> 4$ kpc, and zero inside 4 kpc 
(Scoville \& Sanders 1987).   The molecular-gas surface density has a
a strong peak at the Galactic centre, another peak around 5 kpc from the 
Galactic centre, then decreases outwards (Scoville \& Sanders 1987).
We ignore variations of the surface density in azimuth, and we 
approximate the distribution in $z$ as an exponential, 
$f_0(z)=f_0(0)\exp(-|z|/H)$.  
{}From these surface densities, we compute the filling fraction in the 
mid-plane as,
\begin{equation}
f_0(0)= {\Sigma (\beta-2)(n_1^{1-\beta}-n_2^{1-\beta}) \over
2H \mu (\beta-1)(n_1^{2-\beta}-n_2^{2-\beta})},
\end{equation}
where $\mu =2.72 m_p$ for the GMC, and $\mu = 1.36 m_p$ for the CNM.
At the solar circle, this gives $f_0({\rm GMC})=10^{-3}$, while 
$f_0({\rm CNM})=0.04$.

The warm HI, warm HII, and hot HII we take as constant densities
of 0.3, 0.15, and 0.002 cm$^{-3}$ respectively, mid-plane filling factors of
35, 20, and 40 per cent, as well as scale heights of
0.5, 1, and 3 kpc (Bland-Hawthorn \& Reynolds 2000).  We assume
that the hot HII has a temperature of $10^6$ K, and thus $c_s= 150$ km~s$^{-1}$.
The corresponding surface mass densities are 4, 2, and 
0.2 M$_\odot~$pc$^{-2}$, respectively, which we assume are constant as a
function of Galactic radius.

\section{Accretion rate and luminosity distribution functions}

The distribution function of mass-accretion rate for black holes
located at a given region in the Galaxy is given by
\begin{equation}
{d^2 N \over d\dot M dV}= \int\limits_{M_1}^{M_2} dM 
\int\limits_{n_1}^{n_2} dn 
\int\limits_0^\infty dv {df \over dn} {d^2 n_{\rm BH} \over dM dv }
\delta(\dot M(n,M,v)-\dot M),
\end{equation}
where $dV$ is the volume element and
\begin{equation}
{d^2 n_{\rm BH}\over dM dv} = N_{\rm BH}(R,z) \sqrt{2\over \pi} {v^2 \over \sigma_v^3}
\exp{\left[-{v^2\over 2 \sigma_v^2}\right]} n_M M^{-\gamma}.
\end{equation}
We define $\dot M_0 = \pi G^2 {\rm M}_\odot^2 m_p (1 {\rm km~s^{-1}})^{-3}
= 3.8 \times 10^{14}$, so that $\dot M(n,M,v) = 
\dot M_0 n M^2 (v^2+c_s^2)^{-3/2}$.
For a given mass and sound speed, we can define the minimum number density 
required for accretion at a rate greater than $\dot M$:
$n > n_0 = \dot M c_s^3/(\dot M_0 M^2)$.
Using equations (\ref{bondi}) and (\ref{dndm}), we
can first carry out the $v$ integration analytically,
\begin{equation}\label{dndmdot}
{d^2 N\over d\dot M dV }  = N_{\rm BH} \int\limits_{M_1}^{M_2} dM \, n_M 
M^{-\gamma}\int\limits_{{\rm max}\{n_1,n_0\}}^{{\rm max}\{n_2,n_0\}} dn
f_0 n^{-\beta}
\sqrt{2\over \pi} {v_0 \over \sigma_v^3} 
{(\dot M_0 n M^2)^{2/3}\over \dot M^{5/3}}
\exp{\left[-{v_0^2\over2\sigma_v^2}\right]},
\end{equation}
where $v_0^2=(\dot M_0 n M^2/\dot M)^{2/3} -c_s^2$.  
The remaining two integrals we compute numerically.

\vskip 2mm
\hbox{~}
\centerline{\psfig{file=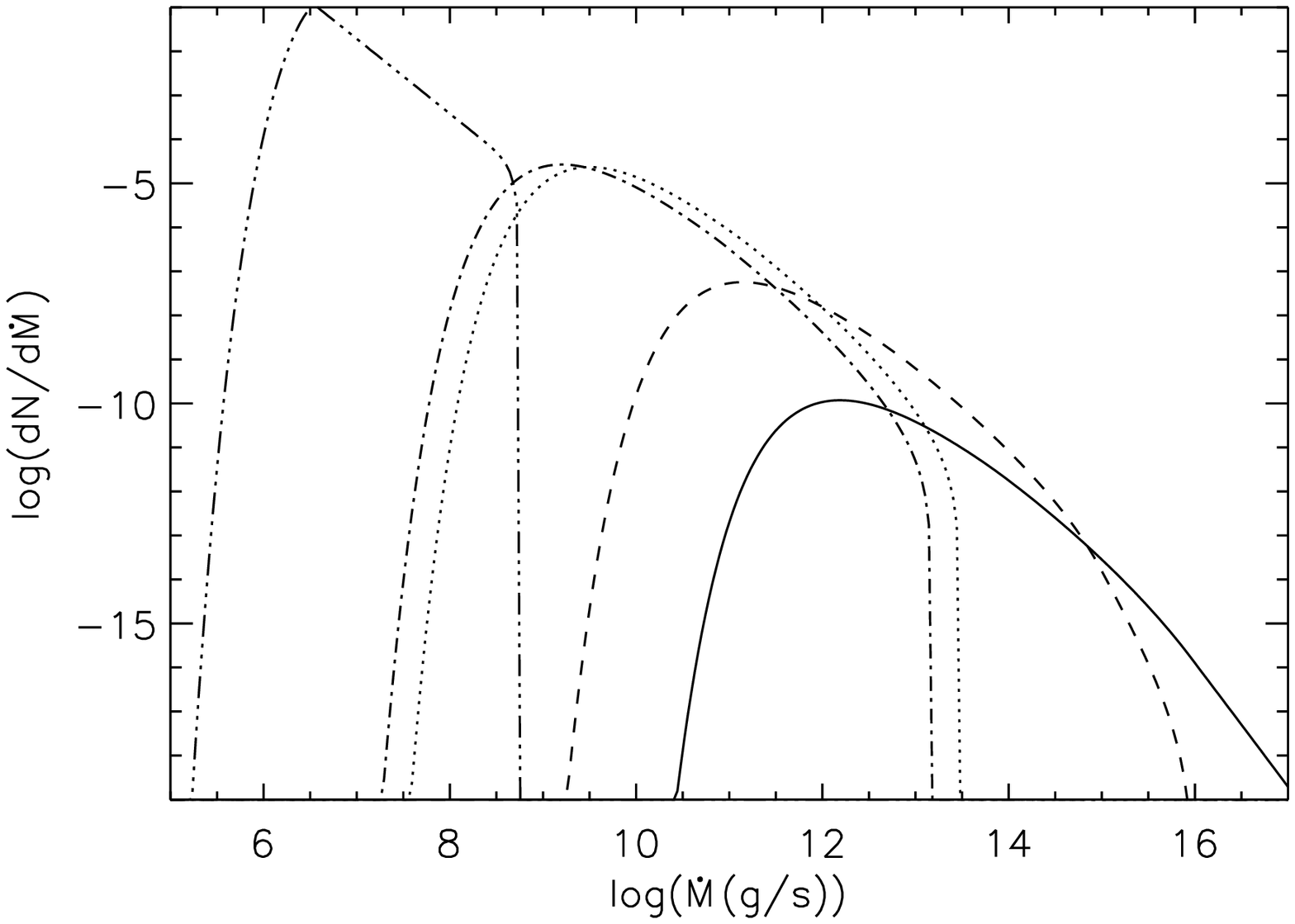,width=5in}} %FIGURE 1
\noindent{
\scriptsize \addtolength{\baselineskip}{-3pt}
\vskip 1mm
\begin{normalsize}
\noindent Fig. 1:  The number density of black holes accreting with
accretion rates between $\dot M$ and $\dot M+d\dot M$ as
a function of $\dot M$ at the solar radius for $\sigma_v=40$ km~s$^{-1}$.
The solid curve refers to black holes accreting
within molecular clouds, while the short dashed curve refers to
black holes within the cold neutral medium.  The dotted curve
is black holes within the warm HI, the dash-dot curve for the
warm HII, while the dash-triple dot curve for the hot HII.
\end{normalsize}
\vskip 3mm
\addtolength{\baselineskip}{3pt}
}

In Figure 1 we plot the function $dN/d\dot M$ for the various phases of the
interstellar medium at the solar circle.   The densest gas (GMCs) dominates 
the highest
accretion rates, while the hottest gas dominates the lowest accretion
rates.  For the hot HII, accretion is subsonic and $n$ is assumed to be
constant, so $\dot M \propto M^2$.  Thus, $dN/d\dot M \propto 
\dot M^{-(1+\gamma)/2}$, spanning a range in $\dot M$ of $(M_2/M_1)^2$, 
or two decades.  This is consistent with the numerical spectrum shown in 
Figure 1.  The other phases have more complicated accretion-rate 
distributions since $\sigma_v > c_s$.

To compute the expected total number of black holes in the Galaxy accreting
at a rate $\dot M$, we integrate the luminosity functions (assumed
to be constant) over the gas filling fraction times the number density
of black holes as a function of position over the entire Galaxy,
\begin{equation} \label{dndmdotgal}
{dN\over d\dot M} = {d^2 N\over d\dot M dV }\Biggr|_{(R_\odot,z=0)} 
\int\limits_0^\infty dR\, 2\pi
R \exp{\left[-{R-R_\odot\over R_{\rm disc}}\right]}
2{\Sigma(R)\over \Sigma(R_\odot)}
\int\limits_0^\infty dz \exp{\left[-{z\over H_{\rm BH}}-{z\over H_{gas}}\right]}.
\end{equation}
Figure 2(a) shows $N(>{\dot M})=\int_{\dot M}^\infty d\dot M' 
(dN/d\dot M')$ for both
black holes and neutron stars.  For neutron stars, we choose $M=1.4 {\rm M}_\odot$, 86 per cent
with $\sigma_v = 175$ km~s$^{-1}$ and 14 per cent with $\sigma_v=700$ km~s$^{-1}$ (Cordes
\& Chernoff 1998),
$N_\odot = 5.2\times 10^5$ kpc$^{-3}$, and $H=1$ kpc, yielding a 
total of $10^9$ neutron stars within our galaxy. 
The total number of accreting black holes is smaller than the total
number in the disk since we have assumed that the gas filling factor decreases 
with scale height.  This reduction may be artificial since gas at some density 
presumably fills the entire Galaxy, but this only affects small accretion rates.

We expect $\sim 10^4 N_9$ isolated black holes to be accreting
at rates comparable to the estimated rates in black-hole X-ray binaries,
$\dot M > 10^{15}$ g~s$^{-1}$ (van Paradijs 1996).  This exceeds
the parent population of black-hole X-ray binaries, which
may be 10$^{2-3}$ (Iben et al. 1995).  
Black holes outnumber neutron stars due to
their larger mass and smaller space velocity as shown in Figure 2(a).  

To compute the number of detectable isolated black holes requires
converting the accretion-rate distribution into a luminosity function.  
Given the uncertainties in accretion physics, we will make two assumptions 
based on what is known about black-hole X-ray binaries: persistent
and transient sources.  The first assumption shown in Figure 2(b) is 
that the sources are time-steady, similar to high-mass 
X-ray binaries such as Cygnus X-1.  Under this assumption, we simply convert
the accretion rate into an X-ray luminosity in a given X-ray band. Shown
is the ionising radiation band from 1--1000 Ryd for 
$\epsilon_{\rm ion}=10^{-5}$;
the horizontal axis simply scales with efficiency.  The black-hole
and neutron-star functions overlap for $\epsilon_{\rm ion} \sim 10^{-5}$
despite the lower black-hole efficiency due to the larger black-hole
mass, smaller velocity
dispersion, and smaller scale height.
The total luminosity of isolated accreting black holes within 
the Milky Way is $\sim 10^{36}N_9\epsilon_{-5}$ erg~s$^{-1}$.

\vskip 2mm
\hbox{~}
\centerline{\psfig{file=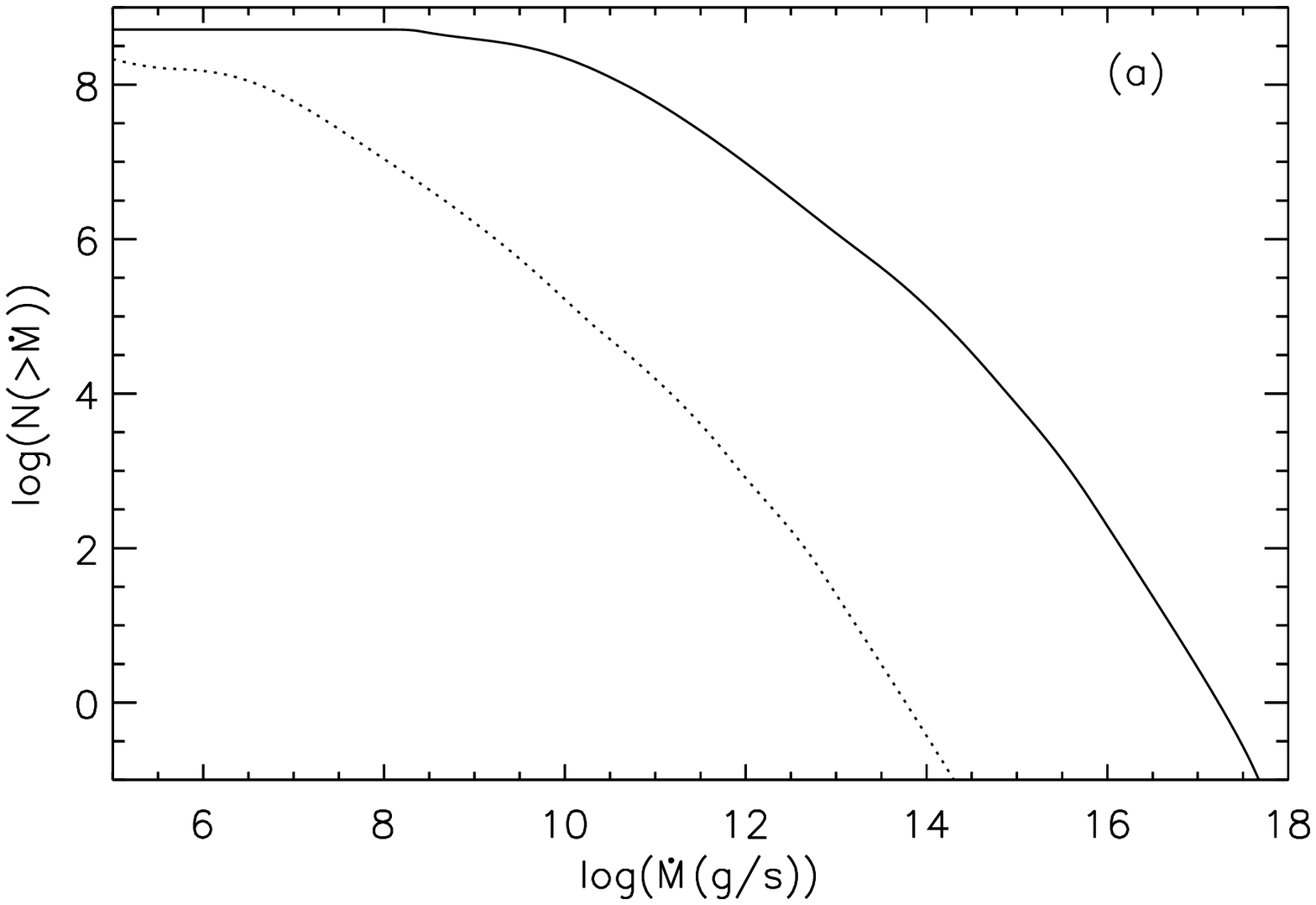,width=5in}} %FIGURE 2
\centerline{\psfig{file=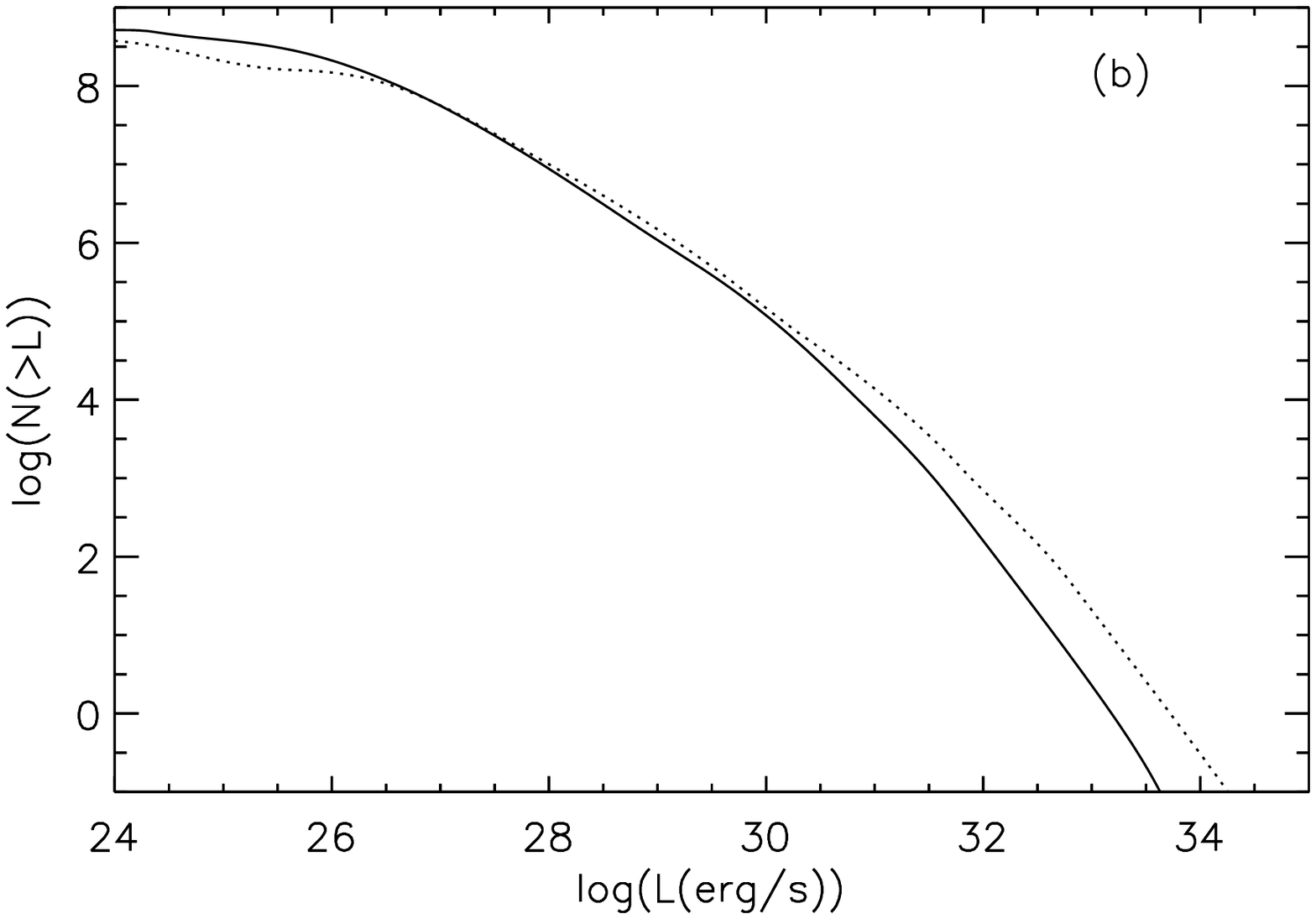,width=5in}} %FIGURE 2
\noindent{
\scriptsize \addtolength{\baselineskip}{-3pt}
\vskip 1mm
\begin{normalsize}
\noindent Fig. 2:  (a) The number of black holes in the Milky Way with
accretion rates greater than $\dot M$ for $\sigma_v=40$ km~s$^{-1}$ (solid line).
For comparison, the dotted line shows the same computation for neutron stars. 
(b) The number of black holes in the Milky Way with luminosities greater 
than $L_{\rm ion}$ with $\epsilon_{\rm ion}=10^{-5}$ and the number of neutron 
stars for $\epsilon_{\rm ion}=10^{-1}$ (dotted line).
\end{normalsize}
\vskip 3mm
\addtolength{\baselineskip}{3pt}
}

The second assumption that the black holes are transient
we discuss in section 6.2 below.

\section{Detection probability}

To compute the detection probability, we need to integrate the number
of sources above a given observed flux as a function of position within
the Galaxy.   We will consider two energy emission bands, `soft' X-ray
{}from 1--10 keV, and `hard' X-ray from 10--100 keV, which have
efficiencies of $\epsilon_{\rm soft}$ and $\epsilon_{\rm hard}$.  Since the
model spectrum we have used is flat, $\nu f_\nu \propto \nu^0$, the luminosity 
per unit decade of photon energy is constant.   Thus, the flux in each
of these bands is $1/3$ of the flux of ionising photons which
covers three decades, or $\epsilon_{\rm soft}=\epsilon_{\rm hard}=\epsilon_{\rm ion}/3$.
A more complex X-ray spectrum will change our
results by factors of order unity,  which we defer to future work given 
the poor theoretical understanding of the spectra of accreting black holes.

For neutron stars, we assume a spectrum composed of a blackbody with $T=0.15$ keV
plus a power law with $\alpha = 1$ (i.e. flat in $\nu F_\nu$) and
the flux from 1--10 keV equal to 1 per cent of the total flux of the blackbody
component.  This was chosen to be similar to the spectra of neutron-star
binaries Cen X-4 and Aquila X-1 in quiescence (Rutledge et al. 2001a,b).
This spectrum yields $\epsilon_{\rm soft}=
0.13 \epsilon_{\rm ion}$ and $\epsilon_{\rm hard}= 0.03 \epsilon_{\rm ion}$ due
to the fact that most of the flux is in the extreme ultraviolet.
We neglect photoelectric absorption since the half-column density of molecular
clouds is approximately $N_H\sim 1\times 10^{22}$cm$^{-2}$, which leads
to absorption of only 36 per cent of the flux in the soft X-ray band
(using the opacities of Morrison \& McCammon 1987).  Toward the
Galactic centre,
the hydrogen column density is about $N_H\sim 6\times 10^{22}$cm$^{-2}$ due
to the ring of gas near the centre, leading to absorption of about 86 per cent of the 
soft flux.  In the hard--X-ray band, the absorption is negligible.

\subsection{Persistent sources}

For the persistent-source assumption, we assume that the accretion flow
emits isotropically, so we simply need to integrate
the luminosity function over the entire galaxy with the
transformation:
\begin{equation}
{dN\over dF dV}  = {dN \over d\dot M dV} {4 \pi D^2 \over 
\epsilon_{\rm soft,hard} c^2},
\end{equation}
where $D$ is the distance to a given point in the Galaxy.
Figure 3 shows the predicted number of sources as a function
of flux integrated over the entire Galaxy for the persistent-source assumption.
Despite the much lower assumed efficiency, the number of detectable black holes
becomes comparable to the number of detectable neutron stars for
$\epsilon_{BH}\sim 10^{-4}\epsilon_{NS}(N_{BH}/N_{NS})^{0.8}$.
The number of black holes above a certain flux at high flux scales approximately
as $N(>F) = 10^4F^{-1.2}$ near $F=10^{-15}\epsilon_{-5}$ erg cm$^{-2}$ s$^{-1}$.
Given the flatness of this dependence,
the best detection strategy is to cover as much area of the sky as possible, 
assuming the constraint of a fixed amount of observing time with an X-ray 
telescope with a sensitivity that scales as $t^{-1/2}$.  Reducing
the mass cutoff from $M_2=50{\rm M}_\odot$ to $M_2=13 {\rm M}_\odot$ only decreases
the number of objects by a factor of $\sim 2$ since there are fewer objects
at high mass (Figure 3).

\vskip 2mm
\hbox{~}
\centerline{\psfig{file=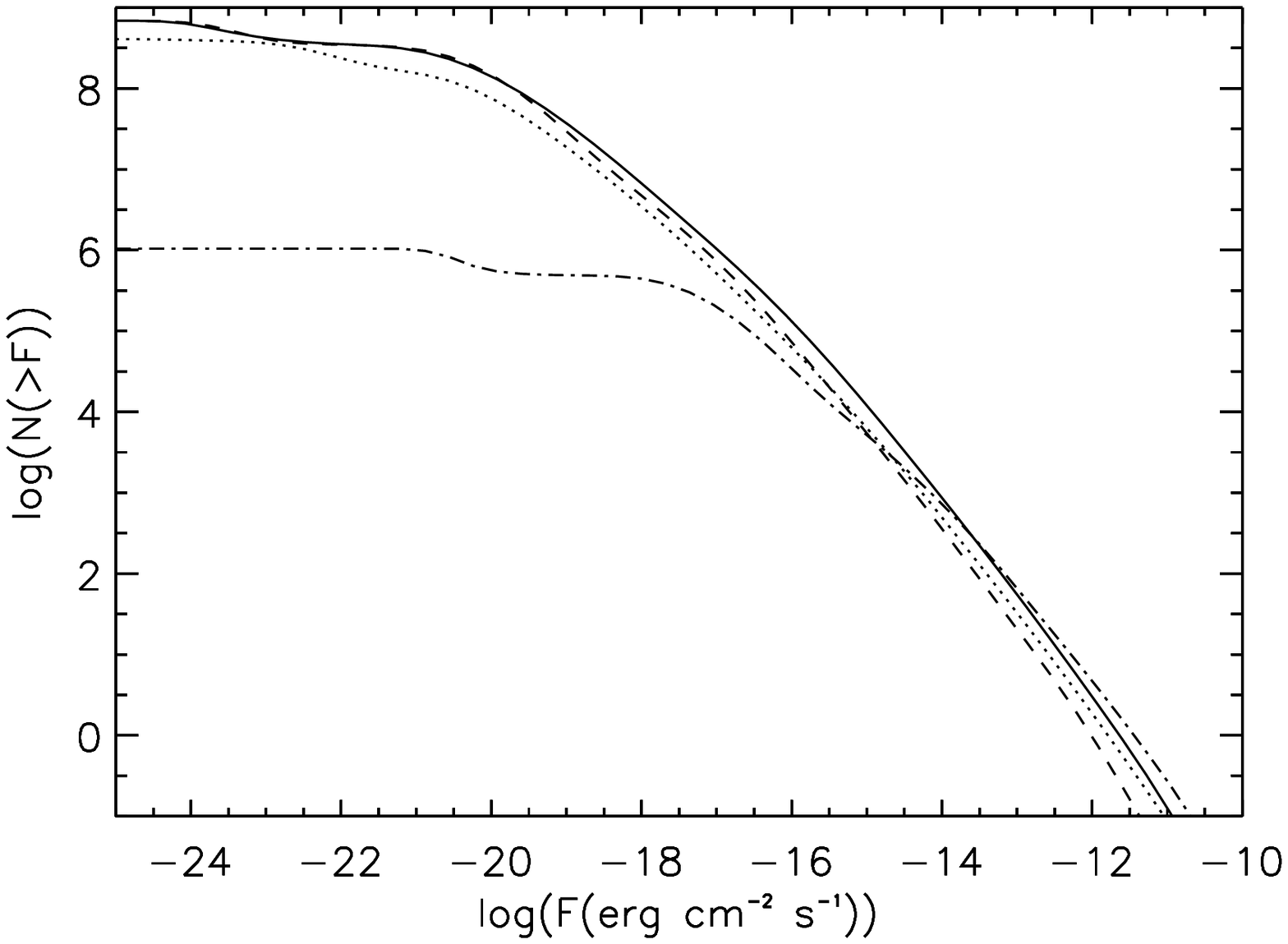,width=5in}} %FIGURE 3
\noindent{
\scriptsize \addtolength{\baselineskip}{-3pt}
\vskip 1mm
\begin{normalsize}
\noindent Fig.~3:  The number of black holes visible to an observer located
within the Solar system as a function of X-ray flux in either the soft
or hard X-ray bands for a persistent accretion efficiency of
$\epsilon_{\rm ion} = 10^{-5}$ with $\epsilon_{\rm soft,hard}
=\epsilon_{\rm ion}/3$ for $M_2=50{\rm M}_\odot$ (solid line) and
$M_2=13 {\rm M}_\odot$ (dashed line).  For comparison the number of
neutron stars above
a given soft X-ray flux is shown as well in the soft X-ray band with
$\epsilon_{\rm soft} = 0.13 \epsilon_{\rm ion} = 0.013$ (dotted line).
In addition, we show predictions for a population of $10^6$ black holes
with mass $250 {\rm M}_\odot$ for efficiencies of $10^{-5}$ (dash-dot line).
Note that the horizontal axis scales with efficiency for each object.
\end{normalsize}
\vskip 3mm
\addtolength{\baselineskip}{3pt}
}

We have also computed the the number density of black holes with fluxes 
greater than $10^{-15}\epsilon_{-5}$ erg~cm$^{-2}$~s$^{-1}$ as a function of Galactic
latitude and longitude for one quadrant of the Galaxy (Figure 4).
The black holes are strongly concentrated toward the Galactic plane
and toward the Galactic centre.  Half of the black holes lie within
the dashed contour which covers an area of 504 deg$^2$ in Figure 4, 
giving an average of 3$N_9$ per square degree within this region near the 
Galactic plane.  We have not attempted to include the inhomogeneity of 
the ISM which might be important for nearby black holes in the local bubble 
or for black holes in interstellar clouds.

\vskip 2mm
\hbox{~}
\centerline{\psfig{file=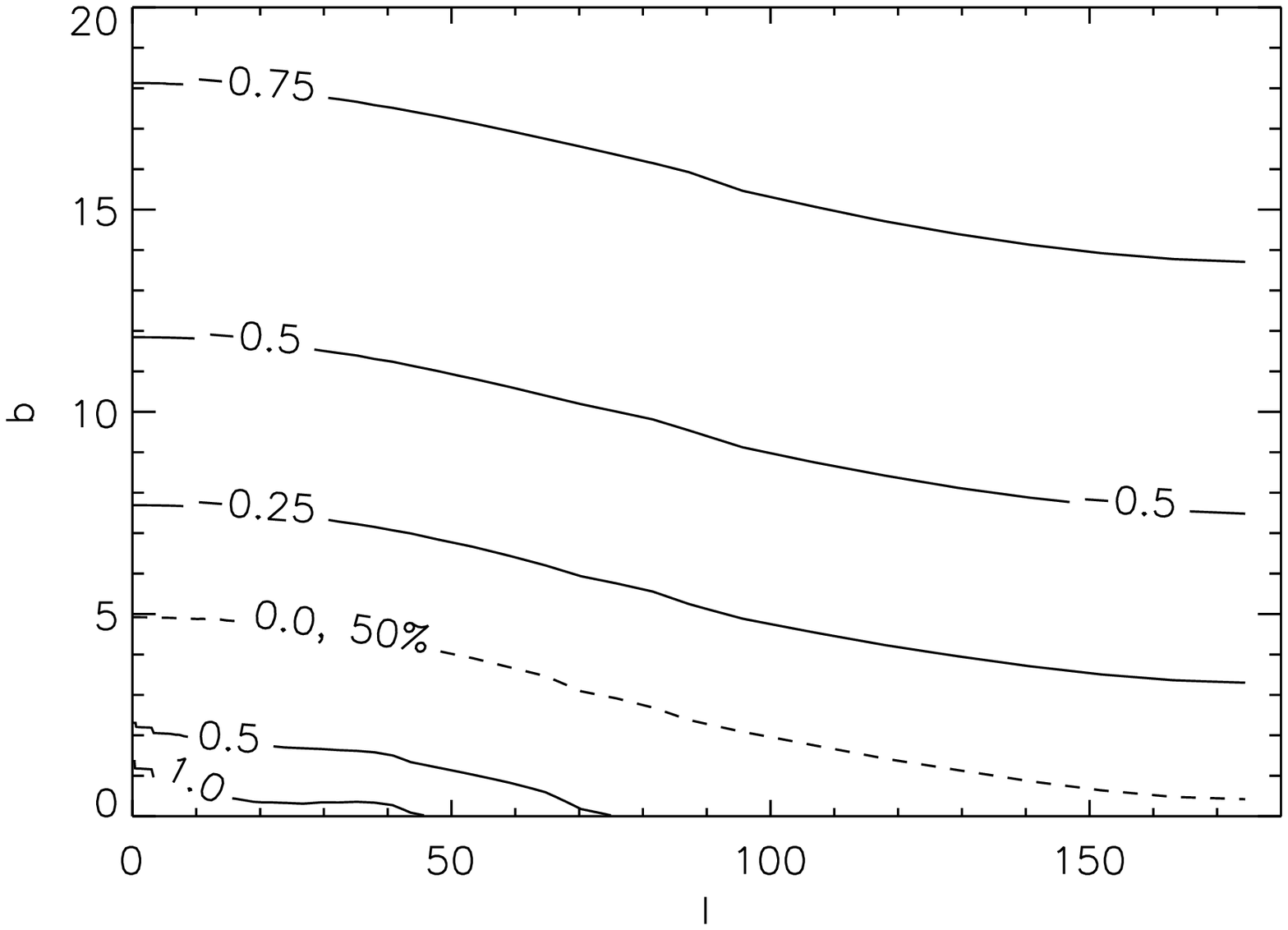,width=5in}} %FIGURE 4
\noindent{
\scriptsize \addtolength{\baselineskip}{-3pt}
\vskip 1mm
\begin{normalsize}
\noindent Fig.~4:  The logarithm of the number density of black holes
per degree square with fluxes above $F_X=10^{-15}\epsilon_{-5}$ erg~cm$^{-2}$~s$^{-1}$ 
in the soft or hard band as a function of galactic longitude, $l(^\circ)$,
and latitude, $b(^\circ)$.
The labels indicate the logarithm of the number density.  The dashed line 
indicates the contour within which 50 per cent of the black holes are contained.
The total number of black holes above this flux level 
is $\sim 5000$, while the maximum number density is 94 per square degree 
at the Galactic Center. 
\end{normalsize}
\vskip 3mm
\addtolength{\baselineskip}{3pt}
}

Finally, we have computed the distribution of sources with distance from
the Sun.   Figure 5 shows the number of black-hole sources with fluxes
greater than $F=10^{-15}\epsilon_{-5}$ erg cm$^{-2}$ s$^{-1}$ less
than a given distance for each of the phases of the ISM.  The total number 
of sources above this flux level is 3800$N_9$ in the GMCs, 6700$N_9$ in 
the CNM, 900$N_9$ in the warm HI, 200$N_9$ in the warm HII, and 0 
in the hot HII.  The average distance for the 
detectable black holes is 5 kpc in the GMCs, 1 kpc in the CNM,
200 pc in the warm HI, and 140 pc in the warm HII.  The average distance
of black holes in all phases is 2 kpc.  Note that there
is a rise in the number of black holes in GMCs at the Galactic centre 
(8 kpc away) since the surface density of molecular gas increases
by a factor of $\sim 10$ within 500 pc of the Galactic centre.  The
sources in the gas clouds tend to be more distant since it
requires looking further in the Galaxy to find clouds, while the
the sources in the interstellar gas tend to be closer since they are
fainter due to lower accretion rates.

The sensitivity of the ROSAT All-Sky Survey (RASS) Bright Source Catalog
is about $10^{-12}$ erg cm$^{-2}$ s$^{-1}$ (Voges et al. 1999; 
we have converted count rate to flux assuming $\alpha=1$), so at most
$3N_9\epsilon_{-5}^{1.2}$ accreting black hole should have been detected 
by this survey given the assumption of persistence.  
We can see immediately from Figure 3 why it may have been difficult to
detect isolated accreting neutron stars with the RASS since we predict
about one visible neutron star on the sky at this flux level, which will
be reduced further by X-ray absorption.  Danner (1998) has surveyed sources
in molecular clouds within the RASS, finding no new isolated neutron-star
candidates, and thus, we presume, no isolated black-hole candidates.  
For an X-ray observatory such as {\em EXIST}, which will reach sensitivities
of $\sim 10^{-13}$ erg cm$^{-2}$ s$^{-1}$ for one-year co-added data
(Grindlay et al. 1999 ), about $\sim 50N_9\epsilon_{-5}^{1.2}$ sources 
may be detectable at hard X-ray energies as well.  {\em EXIST} will have an 
angular resolution of about 2 arcminutes and position sensitivity
of $\sim 30$ arcsec, which should make identification feasible within
molecular clouds.  

The {\em Chandra} and {\em XMM-Newton} observatories have much better sensitivities
$\sim 10^{-14}$ erg cm$^{-2}$ s$^{-1}$ for a 10 kilo-second (ksec) 
observation;  however, both satellites
cover a much smaller field of view ($16'\times 16'$).  The ChaMPlane
survey ({\em Chandra} Multi-wavelength Galactic Plane Survey, Wilkes et al. 2000) 
will cover about 1 square degree per year to a flux of $\sim 10^{-15}$
erg cm$^{-2}$ s$^{-1}$ for the 1--10 keV band.
This flux limit is deep enough that $\sim 10^4N_9\epsilon_{-5}^{1.2}$ sources 
within the Galactic plane might be detectable (Figure 3);  however, the central
50 per cent are spread over an area of 504 deg$^2$ in the Galactic Plane
(Figure 4).  The {\em XMM-Newton}
serendipitous Survey will cover an area of the sky $\sim$ 10 times 
larger than the {\em Chandra} Serendipitous Survey at similar sensitivity 
(Watson et al. 2001).  Thus, these surveys may result in 
$\sim 30N_9\epsilon_{-5}^{1.2} $ detections per year.

One might expect that targeting molecular clouds may be advantageous
for finding black holes.  However,
due to their proximity, the 23 clouds within 1 kpc have angular sizes 
of about 5--20$^\circ$ (Dame et al. 1987), which will make them rather 
difficult to cover with the small field of view of {\em Chandra} or {\em XMM-Newton}.  For
instance, the single pointing in the $\rho$-Ophiucus molecular cloud observed 
by Imanishi et al. (2001) covers a volume $\sim$ 18 pc$^3$
(assuming the cloud is spherical), so we expect only $\sim 3\times 10^{-2}N_9$
black holes to reside within the observed volume.  Observations pointed
at molecular clouds will not improve these detection rates, but observations
pointing toward the Galactic centre should improve the number
of detections significantly.  Surveys targeted at the Galactic Center 
(Baganoff et al. 2001) may find $7 N_9\epsilon_{-5}^{1.2}$ per 
{\em Chandra} or {\em XMM-Newton} field.

\vskip 2mm
\hbox{~}
\centerline{\psfig{file=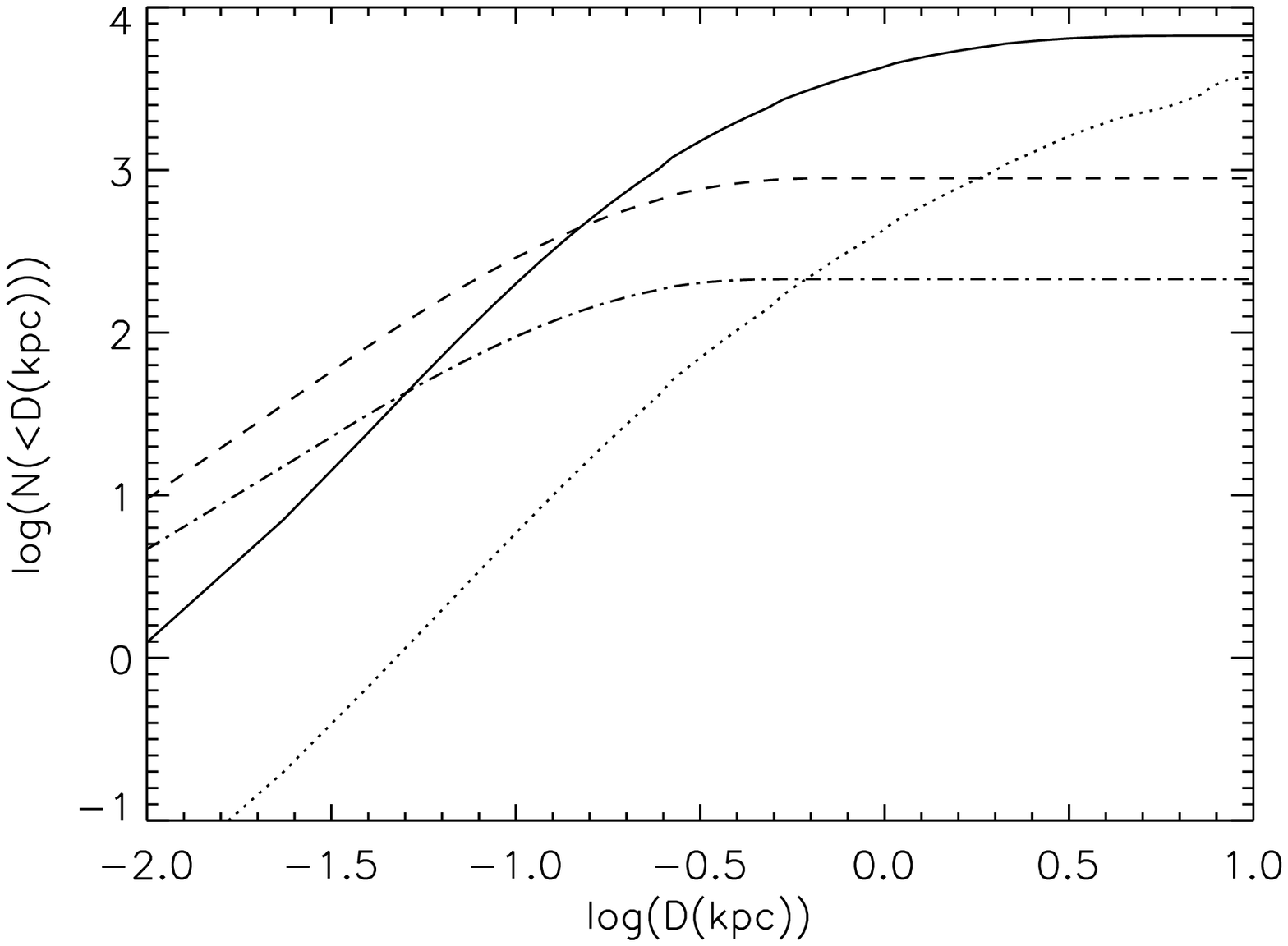,width=5in}} %FIGURE 5
\noindent{
\scriptsize \addtolength{\baselineskip}{-3pt}
\vskip 1mm
\begin{normalsize}
\noindent Fig.~5:  The total number of black hole sources with
fluxes greater than $F=10^{-15}\epsilon_{-5}$ erg cm$^{-2}$ s$^{-1}$ and
less distant than $D$.  The solid line is for the CNM, the
dotted line for the GMCs, the dashed line for the warm HI,
and the dash-dot line for the warm HII.
\end{normalsize}
\vskip 3mm
\addtolength{\baselineskip}{3pt}
}

\subsection{Transient sources}

The second assumption is that each of the sources are transient, driven by the
hydrogen-ionisation instability at radii in the accretion disc with 
$T\sim 5000$ K
as in low-mass X-ray binaries.  In this case, we assume that the sources
undergo outbursts for several months with luminosities near Eddington,
while remaining faint for several decades.
Chen, Shrader \& Livio (1997) have shown that soft X-ray transients
tend to peak at $L_{\rm out}\sim 0.2 L_{\rm Edd}$ with an average duration 
(decay time scale) of
$t_o \sim 20$ days and recurrence times varying from 2 years to 60 years for
black holes.   These are only the observed transients;  transients with
longer recurrence time scales may not have had time to repeat.
If we assume that each black hole accretes quiescently for a time
$t_q$, storing up mass in the accretion disk with a small fraction of
gas accreting on to the black hole, and then all of this
mass is released in an outburst of duration $t_o$, then
\begin{equation}
t_q = 130 yr {L_{\rm out}\over 0.2 L_{\rm Edd}} {t_o\over 20 {\rm days}} 
\left({\epsilon_o \dot M \over 10^{14} {\rm g~s^{-1}}}\right)^{-1}
{M \over 9 {\rm M}_\odot},
\end{equation}
where $\epsilon_o$ is the radiative efficiency of accretion during 
outburst.  The outburst rate in the Milky Way is then
\begin{equation}
\dot N_{\rm out} = \int_{M_1}^{M_2} dM\int_0^\infty d\dot M 
{d^2N\over dM d\dot M} t_q^{-1}.
\end{equation}
Assuming a 10 per cent efficiency during outburst, we find 
$\dot N_{\rm out} \sim 536N_9$ yr$^{-1}$ for $\sigma_v=40$ km~s$^{-1}$.
This is clearly inconsistent with observations, demonstrating
that not all isolated black holes undergo transient
outbursts if the assumptions of our calculation are correct.
If only black holes accreting above $\dot M > 10^{15}$ g~s$^{-1}$
undergo outbursts, consistent with the observed X-ray novae
(van Paradijs 1996), then these numbers reduce to 50 yr$^{-1}$,
still large given that about $\sim 1$ transient is detected
each year, and generally they are found to have companions.  
We conclude that only a small fraction of isolated black holes
might experience transience similar to X-ray novae.

\subsection{Detecting microlensing candidates}

The uncertainty in distance of the MACHO black-hole microlensing candidates
leads to an uncertainty in their mass, and thus an uncertainty in their
status as black holes.  One possible way to constrain the mass of these
lenses is to look for X-rays from accretion.  Figure 6 shows the flux as a 
function of distance for MACHO 96-BLG-5, MACHO 98-BLG-6, and OGLE-1999-BUL-32 
(Mao et al. 2001) for a number density of 1 cm$^{-3}$ 
with the total velocity set to $\sqrt{3/2}$ times the observed sky velocity 
projected on to the lens plane (ignoring the unknown projected source velocity)
and subtracted from the Galactic rotation velocity
(220 km~s$^{-1}$) and an X-ray efficiency of $\epsilon_{\rm ion}=10^{-3}$.
We note that if the sky velocity of the source is significant, then the
inferred lens velocity will be different, typically smaller, which will
increase the luminosity.  The flux scales as $F\propto \epsilon n D^{-4}$ 
for small distances since the mass depends on the inverse of the distance 
and the flux decreases as the inverse of distance squared.   
Observing in the X-ray would be advantageous to avoid confusion with the 
background microlensed star.  In the soft X-ray, a 1--10 keV flux of 
$\sim 10^{-15}$ erg cm$^{-2}$ s$^{-1}$ can
be detected with a $\sim$ 400 ksec exposure with {\em Chandra} or $\sim$ 200 ksec
with {\em XMM-Newton}.  A detection of a black-hole candidate at such a flux level will allow
an upper limit on the distance of $\sim 1-3$ kpc to be placed, assuming they
accrete from the diffuse ISM with $n < n_{\rm max} =1$ cm$^{-3}$ and assuming 
an X-ray efficiency less than $\epsilon_{\rm max} =10^{-3}$.  A decrease in 
efficiency will lead to a decrease in flux, and thus a decrease in the distance 
upper limit.  An upper limit on distance converts into a lower limit on mass, 
since $M \propto D^{-1}$, which may confirm the black-hole hypothesis.  The 
mass lower limit is proportional to $M_{\rm min} \propto (\epsilon_{\rm max} 
n_{\rm max}/F)^{-1/4}$, a weak dependence on our assumed parameters.
A cloud or companion
could increase the accretion rate, resulting in a higher luminosity, thus
a larger distance for the same flux; however, clouds occupy less than 5 per cent
of the volume near the Galactic plane and can be searched for 
in HI 21 cm or CO absorption or emission (the MACHO fields are toward
Baade's window, thus unlikely to contain a high molecular column density), 
while companions should be detectable with HST.

\vskip 2mm
\hbox{~}
\centerline{\psfig{file=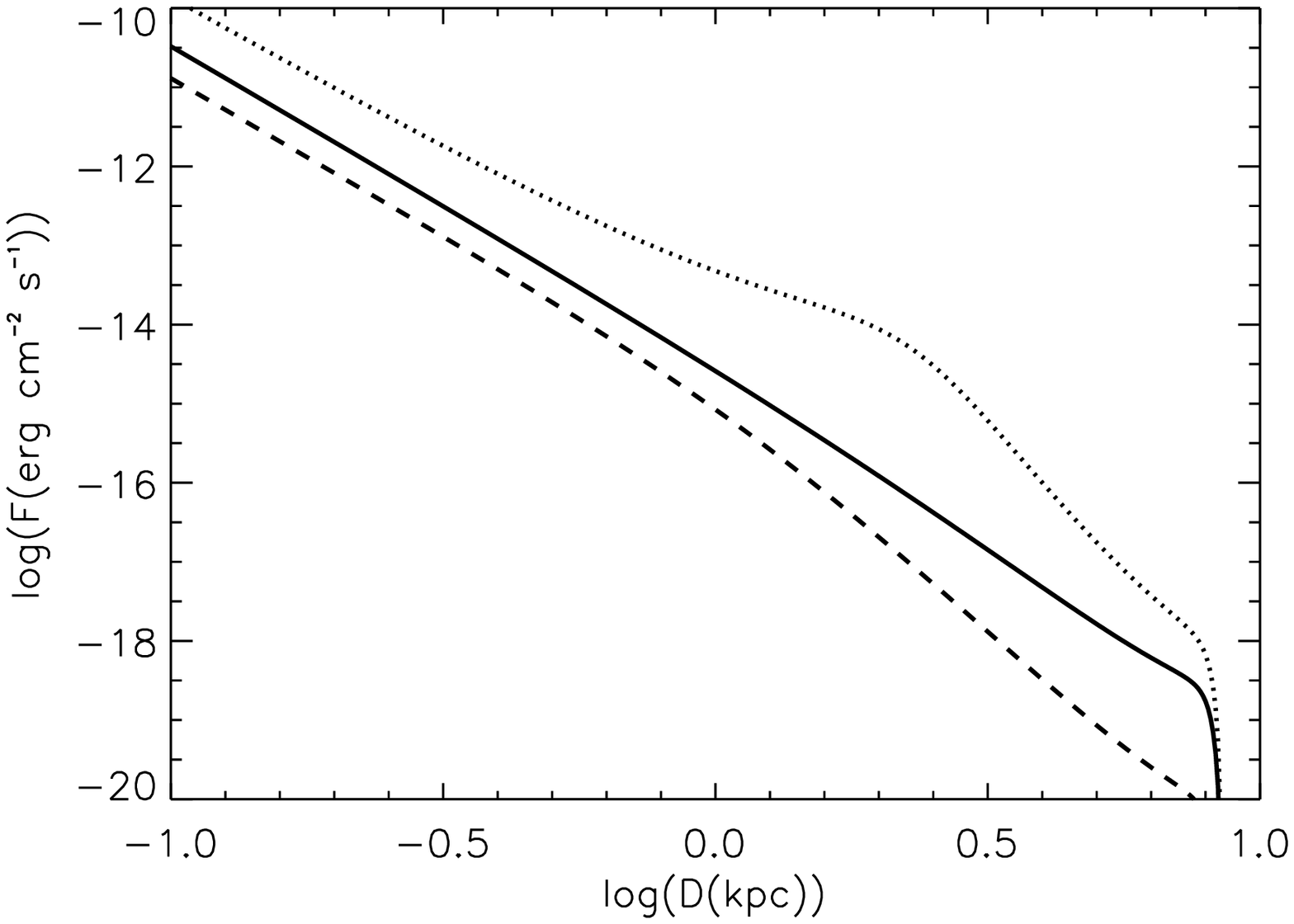,width=5in}} %FIGURE 6
\noindent{
\scriptsize \addtolength{\baselineskip}{-3pt}
\vskip 1mm
\begin{normalsize}
\noindent Fig.~6:  The flux as a function of distance for MACHO 96-BLG-5 
(solid), MACHO 98-BLG-6 (dashed) and OGLE-1999-BUL-32 (dotted) as a function 
of distance for $\epsilon_{\rm ion}=10^{-3}$.
\end{normalsize}
\vskip 3mm
\addtolength{\baselineskip}{3pt}
}

\section{Conclusions}

We have estimated the number of isolated black holes that might be revealed
by their X-ray emission from accretion of interstellar gas.  
We have improved upon previous calculations by taking into account 
the density distribution of the interstellar medium, 
as well as using the known properties of black-hole X-ray binaries and MACHO
black-hole candidates to constrain the phase space and accretion efficiency 
of isolated accreting black holes.  We conclude that persistent isolated black 
holes may be competitive with neutron stars in creating detectable X-ray flux
if their efficiencies differ by $\epsilon_{BH} \sim 10^{-4} \epsilon_{NS}
(N_{BH}/N_{NS})^{0.8}$. 
The ROSAT survey did not have the sensitivity to
detect either isolated accreting neutron stars or black holes;  
an all-sky survey with two orders of magnitude more sensitivity,
$\sim 10^{-14}$ erg~cm$^{-2}$~s$^{-1}$~dex${-1}$, may be able to detect 
$\sim 10^3N_9\epsilon_{-5}^{1.2}$ accreting black holes.
The X-ray observatories {\em XMM-Newton} and {\em Chandra} may be able 
to detect this population with pointed observations; however, they suffer
{}from having small fields of view.  Despite this limitation, Galactic plane
surveys being carried out with these telescopes may be able to detect
tens per year for $\epsilon_{-5}=1$.  If the accreting sources with 
$\dot M > 10^{15}$ g~s$^{-1}$ go through transient outbursts, then 50 isolated 
black holes per year would be detectable as X-ray novae, so we reject
this possibility since isolated X-ray novae have not been observed.

The greatest uncertainty in our predictions is the efficiency of black-hole 
accretion, with which we have parameterized our results.
Most of the detectable black
holes should reside in interstellar clouds that have higher densities;
however, this leads to the problem of confusion with other X-ray sources
such as the coronae of massive stars.   Some tests for whether an accreting
object is a black hole are the following:  1)  Is there high-energy emission?
Accreting black holes tend to show spectra which have power laws extending up 
to $\sim 10^2$ keV (Grove et al. 1998).  2)  What is the nature of the 
variability?  Accreting black holes show no pulsations, show only 
QPOs with $\nu < 1$ kHz, and show power spectra that cut off around 500 Hz 
(Sunyaev \& Revnivtsev 2000).  3)  What is the mass?  Without a binary 
companion, the mass of an accreting object is difficult to measure;  
however, this might be achieved by carrying out astrometry of background 
stars to look for gravitational distortion by the accreting object 
(Paczy\'nski 2001).  4)  What do other parts of the spectra look like?  
Accreting black holes can produce relativistic radio jets (Mirabel \& 
Rodr\'iguez 1999), and photoionisation of the surrounding gas might result 
in observable infrared lines (Maloney, Colgan \& Hollenbach 1997).

There are several assumptions in our calculation that might affect the
results.  We have assumed that the accretion flow is one-dimensional,
and that the sonic point is at the accretion radius.  We have also assumed
that the accretion is time steady;  this may not be the case for higher
accretion rates than we find for black holes accreting from the ISM
(e.g. Grindlay 1978).  If the black hole is moving slower than the sound 
speed at the accretion radius then the preheated region outside the 
accretion radius may have a chance to expand into the ISM, reducing the 
number density, and thus reducing the accretion rate.  We have neglected 
mechanical feedback, which may occur if a jet or wind is developed.  
Recent work on non-radiating
accretion flows indicates that strong winds can be formed which carry the
bulk of the energy outwards as mechanical rather than radiative energy
(Blandford \& Begelman 1999, Igumenshchev, Abramowicz \& Narayan 2000, 
Hawley, Balbus \& Stone 2001).  These authors argue that the accretion
rate scales as $\dot M \propto r$.  If the outer radius is taken
as $r_A$, then $\dot M (10 r_g) \sim 10 (v/c)^2 \dot M(r_A) \sim 2\times 
10^{-7} v_{40}^{-2} \dot M(r_A)$.  If the outer radius is taken to be the
circularization radius, then $\dot M(10 r_g) \sim 10^{-3}
(M/9{\rm M}_\odot)^{-2/3} [(v^2+c_s^2)^{1/2}/40 {\rm km~s^{-1}}]^{-10/3}$.
Either of these circumstances will make black holes less visible,
which in the formalism of this paper corresponds to a further reduction in 
the accretion efficiency.
Magnetic fields will likely play an important role as they are amplified
by flux-freezing, possibly heating the gas to virial temperatures 
(Igumenshchev \& Narayan 2001) and can transport angular momentum via 
the magnetorotational instability once the circularization radius is 
reached, modifying the dynamics of the accretion flow.
We have not explored the dependence of the efficiency on other parameters,
which may result, for example, from changes in the gas density and
angular momentum as a function of the accretion rate, so our extrapolation
{}from estimated efficiencies of black-hole X-ray binaries may be too
optimistic.  The validity of these assumptions can be tested with 3-D 
radiation MHD simulations of a black hole accreting from an inhomogeneous 
medium, a daunting numerical problem.

It is possible that a population of intermediate-mass black 
holes (IMBHs) exist with larger masses around $250 {\rm M}_\odot$, the remnants
of the first generation of star formation (Madau \& Rees 2001).  The larger
masses of these objects would result in yet larger accretion rates and
luminosities by a factor of $\sim 10^3$, if they are distributed with
same phase-space distribution as $9 {\rm M}_\odot$ black holes. If we assume that 
$\sim 10^6$ IMBHs reside in our galaxy, then we find that the number of 
detectable objects at high fluxes may comparable to $9 {\rm M}_\odot$ 
black holes accreting at the same efficiency (Figure 3).  
However, if these objects reside in the halo or bulge of the galaxy,
the number detectable will be decreased significantly.

\section*{ACKNOWLEDGMENTS}

We acknowledge D.~Bennett, L.~Bildsten, O.~Blaes, R.~Blandford, 
J.~Carpenter, 
D.~Chernoff, G.~Dubus, A.~Esin, C.~Fryer, J.~Grindlay, 
T.~Kallman, L.~Koopmans, J.~Krolik, Y.~Lithwick, A.~Melatos, 
S.~Phinney, M.~Rees, R.~Rutledge, N.~Scoville, and K.~Sheth 
for useful conversations and ideas which greatly improved this work.  This 
work was supported in part by NSF AST-0096023, NASA NAG5-8506, and DoE 
DE-FG03-92-ER40701.  Support for the work done by EA was provided by the
National Aeronautics and Space Administration through Chandra
Postdoctoral Fellowship Award Number PF0-10013 issued by the Chandra
X-ray Observatory Center, which is operated by the Smithsonian 
Astrophysical Observatory for and on behalf of the National Aeronautics
Space Administration under contract NAS8-39073.

\bsp
\label{lastpage}
\end{document}